\documentclass[11pt]{article}
\usepackage{geometry}                % See geometry.pdf to learn the layout options. There are lots.
\usepackage{graphicx}
\usepackage{amssymb}
\usepackage{epstopdf}
\usepackage{caption}
\usepackage{threeparttable}
\usepackage{array}

\usepackage[colorlinks = true,
            linkcolor = blue,
            urlcolor  = blue,
            citecolor = blue,
            anchorcolor = blue,
            hypertexnames=false] %JLO% -- Remove this line
            {hyperref}

\def\authorlist#1#2{
    \vskip 0.4in
\begin{center}\begin{large} {\bf  #1 } \end{large}
    \vskip 0.2in
              #2
     \vskip 0.2in
   \end{center}
}

\begin{document}

\title{\textbf{Snowmass 2021 \\ Underground Facilities and Infrastructure Overview \\ Topical Report}}

\maketitle

\authorlist{L.~Baudis, J.~Hall, K.T.~Lesko, J.L.~Orrell}{}

\tableofcontents

%JLO-arXiv%\setcounter{chapter}{5} 

%% IMPORTANT:   from this file, refer to the bibliography as Underground/UF06/bibliography.tex   
%%    refer to a figure   A.pdf  as   Underground/UF06/figures/A.pdf  .

%JLO-arXiv%\chapter{Underground Facilities and Infrastructure Overview}

%JLO-arXiv%\authorlist{L.~Baudis, J.~Hall, K.T.~Lesko, J.L.~Orrell}{}

%\textcolor{red}{Link to original document (remove this link!):} \\
%\url{https://docs.google.com/document/d/1nqeESVXGWopXnaxBdI0xZygkYaFNKE8F7xnDG-ehuz8/edit}

\section{Introduction}
The Underground Frontier (UF) was charged with assessing the anticipated needs and available space for well shielded underground space to conduct scientific research.  The primary focus of this assessment was on:
\begin{itemize}
    \item High Energy Physics research including traditional disciplines of neutrinos, dark matter, and double beta decay;
    \item physics research in other physics fields including solar neutrinos; enabling technologies such as low background assay, background control, and ultrapure materials; 
    \item synergistic research in biology, geology, and engineering; and 
finally identify emergent fields.
\end{itemize}
UF assessed the requirements and needs for these efforts including such inputs as depth, radioactivity control, power, space, access, and occupancy. These are presented in the previous four chapters. Finally the Frontier sought to understand the existing space, space that would become available in the coming decade, and potential for creating additional space to host these efforts. To this end UF created a survey which was sent to existing underground laboratories (see below). This assessment was necessary for UF to develop a bold plan to foster and empower broad reaching underground research, maximize cost savings and synergistic research opportunities, and advance the HEP research program requiring underground space for the next two decades.

\section{Underground Laboratories}
The first well-known use of the deep underground for physics research was the Davis Solar Neutrino Experiment (Nobel Prize 2002) hosted in the then active Homestake Gold Mine in South Dakota. This was followed by Baksan (1977), IMB (1979), Kamioka (1982) - Nobel Prizes 2002 \& 2015, Modane (1982), Gran Sasso (1985), SNO (1999) - Nobel Prize 2015, Boulby (2001), Canfranc (2006), SNOLAB (2009), Jinping (2011, expanded in 2014), SURF (2012, expanded by 2025), Y2L(2003, expanded in 2014), and most recently by Yemilab (2022). The world wide status of underground laboratories is succintly captured in Fig.~\ref{fig:lab-depth}.
\begin{figure}[ht!]
    \centering
    \includegraphics[width=0.9\textwidth]{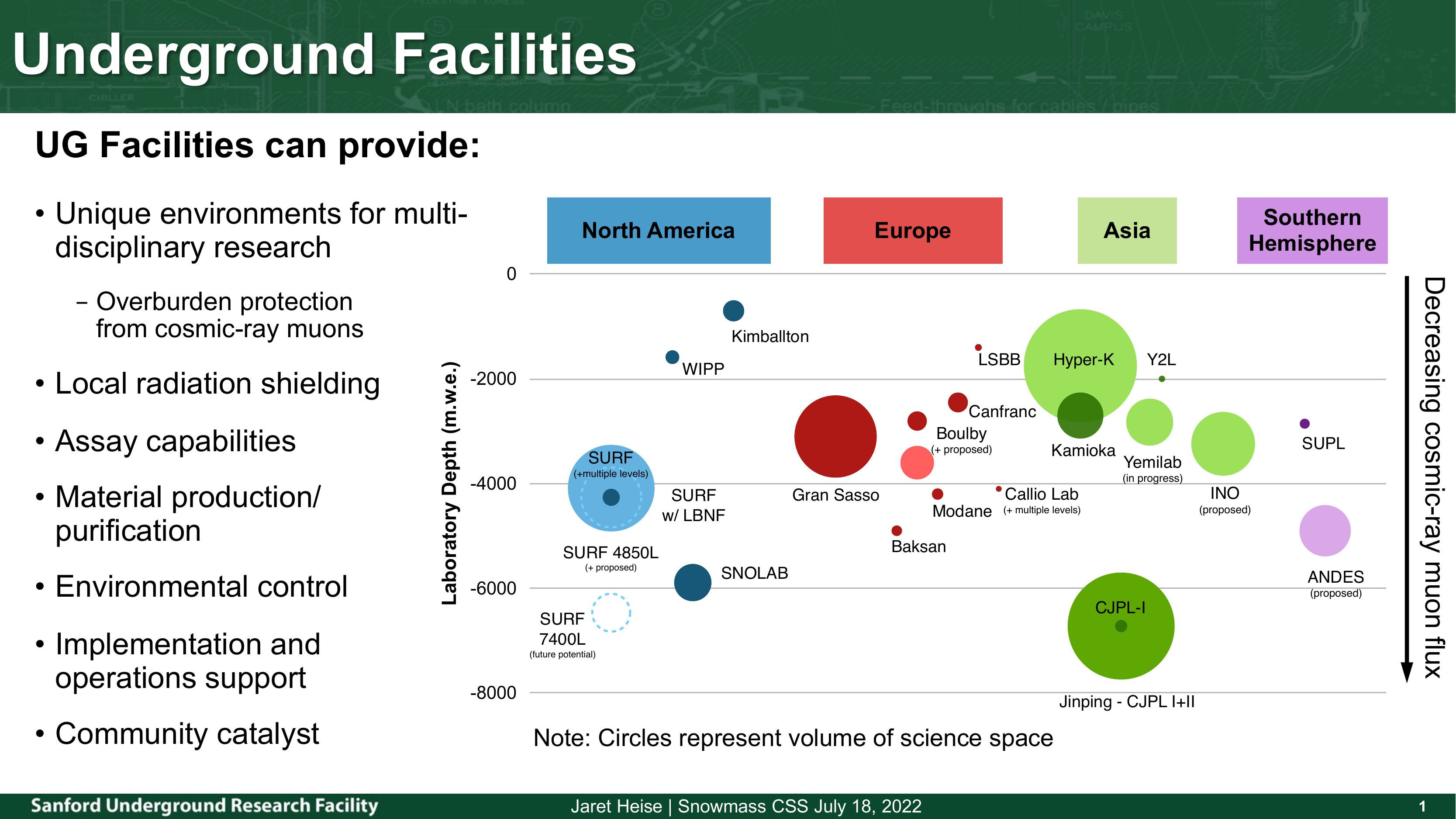}
    \caption{Overview of underground facilities located around the world, showing various key characteristics in depth and underground laboratory capacity. Slide from Jaret Heise (SURF).}.
    \label{fig:lab-depth}
\end{figure}

\section{Underground Science}
The motivation to conduct physics experiments underground is primarily to obtain a well-shielded, stable environment: shielded from backgrounds including cosmic rays and normally occurring radioactivity. This special environment is required by rare search experiments, in which signals of interest can number a few per year per tonne of detector medium.  Experiments involving neutrinos, high energy cosmic rays, nuclear astrophysics, and exotic searches including dark matter, nucleon decay, supernovae neutrino detection. More recently this has been expanded to include gravity waves and violations of fundamental symmetries.  

Other disciplines using underground space include:
\begin{itemize}
    \item geomicrobiology – investigating life in extreme conditions, the origins of life, and new forms of life;
    \item geophysics – investigating coupled processes, resource identification, earthquakes, and seeking to develop advance models of the earth;
    \item geoengineering – striving to understand rock properties in a variety of environments and address societal needs. 
\end{itemize}

In the past twenty years APS, NRC, NAS, and multidisciplinary studies have addressed these experiments comprehensively. In this Snowmass we have highlighted several new users of the underground – quantum information science (QIS), quantum computing (QC), and atom interferometry. 

\subsection{Advances Since 2014 P5 report}
The past ten years have witnessed remarkable progress on a variety of high profile physics topics addressed in the 2014 P5 report.

Second generation direct detection dark matter experiments have been funded.  LZ and ADMX are fully operational and collecting data. LZ, hosted at SURF, recently set world-leading limits on excluding WIMP-like dark matter. ADMX, hosted at the University of Washington,  collected an impressive quantity of data before undergoing upgrades to enhance it sensitivity.  SuperCDMS at SNOLAB underwent rebaselining to address cost issues and is again advancing integration efforts in Canada. Roughly speaking these experiments will operate through this decade.

\begin{figure}[ht!]
    \centering
    \includegraphics[width=0.9\textwidth]{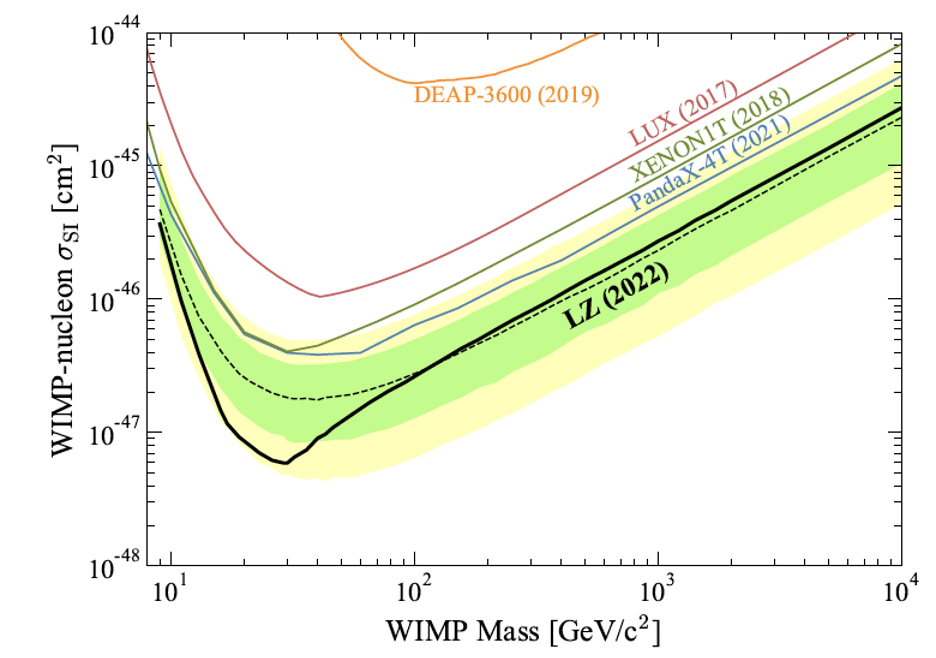}
    \caption{Particle dark matter limits from the LZ experiment~\cite{https://doi.org/10.48550/arxiv.2207.03764}. The 90\% confidence limit (black line) for the spin-independent WIMP cross section vs. WIMP mass. The green and yellow bands are the 1$\sigma$ and 2$\sigma$ sensitivity bands. The dotted line shows the median of the sensitivity projection. Also shown are the PandaX-4T, XENON1T, LUX, and DEAP-3600 limits.}
    \label{fig:lab-depth}
\end{figure}

The past decade has seen significant attention and activity in pursuing new technologies to develop sensitivity, in particular for lighter dark matter candidates. Responding to DOE’s Basic Research Needs for Dark-Matter Small Projects New Initiatives led to the pursuit of three experiments requiring underground locations: Oscura and TESSERACT. These technologies using a variety of techniques and media promise to provide revolutionary sensitivity to dark matter from $\sim$ev to the traditional WIMP scale.

Very remarkable success in underground science has been in defining neutrino masses and mixing.  All three mixing angles have been measured now using a variety of reactor, accelerator, and atmospheric neutrino sources.  The outstanding questions of CP violation and the presence of additional sterile neutrino species remain.  The former is being addressed in the US with the LBNF/DUNE program and HyperKamiokande in Japan. Each of these are very major construction projects.

Limits on neutrinoless double-beta decay have also significantly reduced the allow parameter space for discovery of Majorana neutrinos. Recently the DOE Office Nuclear Physics addressed the U.S. efforts in this field and reviewed efforts using Germanium (LEGEND), Xenon (nEXO), and Molybdenum (CUPID). CUPID is based at LNGS, nEXO is pursuing SNOLAB for a host, various phases of LEGEND are pursuing SNOLAB and LNGS.  It is notable that several Generation 3 Dark Matter experiments have significant sensitivity for neutrinoless double-beta decay, c.f. XLZD white paper.

\section{Underground Laboratory Survey}

As an outcome of the Snowmass process the UF created a survey to assess the characteristics, resources, current occupants, current science program, expansion plans, and anticipated availability in the coming decade. The categories in the questionnaire are listed below.

The questionnaire was sent to major underground laboratories likely to host US participants: SNOLAB, Yemilab, Modane, SURF, Gran Sasso (LNGS), KURF, Canfranc, Boulby, Kamioka,
Reminders were periodically sent if a response was not obtained.  Dialog with several facilities clarified the requested input.  Ultimately questionnaire responses were received from SURF, SNOLAB, Yemilab, Boulby, Kamioka with varying degrees of specificity in their responses. 

\newpage

\begin{center}

    \textbf{Underground Laboratory Survey Questionnaire}
    
    \textit{Categories and details requested from underground laboratories}

\end{center}

{
\small

\begin{minipage}[t]{.5\textwidth}
\raggedright
\begin{itemize}
    \setlength\itemsep{-0.25em}
    \item \textbf{Agency, Authority, Bureaucracy}
    \begin{itemize}
       \setlength\itemsep{-0.25em}
        \item Operations entity
        \item Operations contact
        \item Oversight Agencies and Regulatory Agencies
        \item Authority Having Jurisdiction (life safety)
        \item Facility or access shared with other activities or entity
        \item Training and Security Requirements
    \end{itemize}
    \item \textbf{Environment}
    \begin{itemize}
        \setlength\itemsep{-0.25em}
        \item Depth and Shielding [mwe]
        \item Ambient Temperature, min/max [$^o$C]
        \item Humidity, min/max [\%]
        \item Wall finish and ground support
        \begin{itemize}
            \setlength\itemsep{-0.25em}
            \item Rn concentration [Bq/m$^3$]
            \item muon flux 
            \item muon mean energy
            \item neutron flux and spectrum
            \item gamma-ray flux and spectrum
        \end{itemize}
    \end{itemize}
    \item \textbf{Space(s)} (Each significant space, by category\ldots)
    \begin{itemize}
        \setlength\itemsep{-0.25em}
        \item Currently subscribed
        \item Available
        \item Becoming available in the next 10 years
        \item Projected or planned expansion (new space)
        \item Floor loading capacity [kPa]
    \end{itemize}
    \item \textbf{Utilities (available, not committed)}
    \begin{itemize}
        \setlength\itemsep{-0.25em}
        \item Electrical power [kW]
        \item Standby Power [kW]
        \item Chilled water [kW]
        \item Waste heat to air [kW]
        \item Purified water [m$^3$]
        \item Potable water [lpm]
        \item Compressed air [SCFM]
        \item Network [GB/s]
    \end{itemize}
\end{itemize}
\end{minipage}
\hfill
\noindent
\begin{minipage}[t]{.5\textwidth}
\raggedleft
\begin{itemize}
    \setlength\itemsep{-0.25em}
    \item \textbf{Access}
    \begin{itemize}
       \setlength\itemsep{-0.25em}
        \item Vertical
        \item Horizontal
        \item Facility lockout/maintenance/restricted access periods
        \item Shifts/day access or Multiple-shift access
    \end{itemize}
    \item \textbf{Occupancy}
    \begin{itemize}
        \setlength\itemsep{-0.25em}
        \item Peak Installation Occupancy [persons/shifts]
        \item Steady-state Occupancy [persons/shift]
        \item Underground Refuge Capacity and Duration [persons/days]
    \end{itemize}
    \item \textbf{Underground Assembly Support}
    \begin{itemize}
        \setlength\itemsep{-0.25em}
        \item Cleanroom size and class
        \item Rn-reduced air
        \item Crane [tonne]
        \item Rail/wheeled [dimension-m, capacity-tonne]
        %\item Low Background Assay Facilities (type and number)
    \end{itemize}
    \item \textbf{Surface Facilities}
    \begin{itemize}
        \setlength\itemsep{-0.25em}
        \item Office, Meeting, Control rooms
        \item Cleanrooms
        \item Assembly shop
        \item Machine shop
        \item Electronics shop
        \item Storage/warehouse
        \item Chemistry Laboratories
       %\item Low Back Assay (ICPMS,  alpha, beta, gamma, screening)
    \end{itemize}
    \item \textbf{Technical and Operations Staff}
    \begin{itemize}
        \setlength\itemsep{-0.25em}
        \item Technical Staff
        \item Scientific Staff
        \item EH\&S Staff
        \item Administrative Staff
    \end{itemize}
\end{itemize}
\end{minipage}

}

\newpage

\section{Responses to Survey}

%
% SURF
%
\subsection{Sanford Underground Research Facility (US)}
\vspace{-0.45em}

\begin{figure}[ht!]
    \centering
    \includegraphics[width=0.75\textwidth]{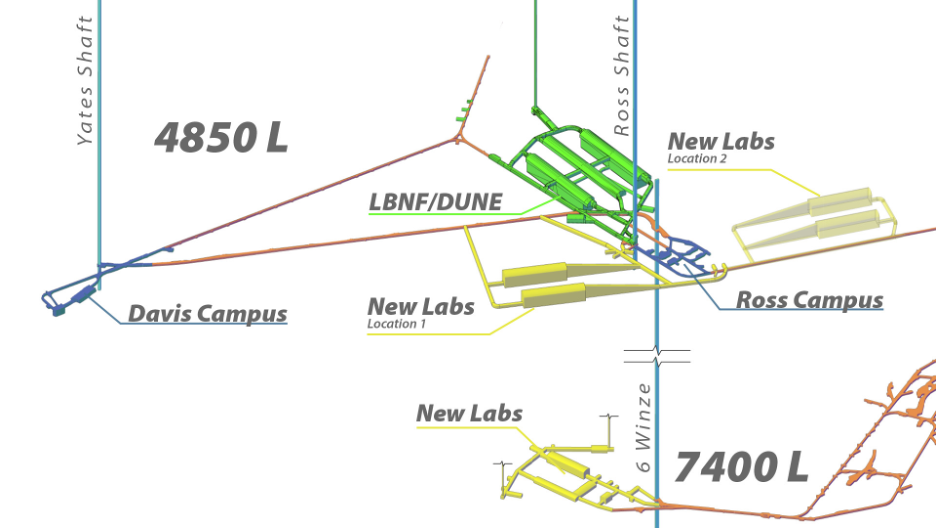}
    \caption{SURF with Expansion Lab Modules.}
    \label{fig:SURF}
\end{figure}

\paragraph{Description:}
\begin{itemize}
\vspace{-0.45em}
\setlength\itemsep{-0.25em}
    \item Former Homestake Gold Mine, multiple drifts and access -- surface to 4850$^{\prime}$ (8000$^{\prime}$)
    \item Past Experiments -- Davis Solar Neutrino Experiment, LUX, {\sc Majorana Demonstrator}
    \item Current Program -- LZ (DM), MJ-$^{180}$Ta, CASPAR (Nuclear Astrophysics), BHUC (Low Background Assay), Geomicrobiology, Geoengineering, Education \& Outreach
    \item Under Construction -- LBNE/DUNE (DUNE excavations $\sim$35\% complete, FY24)
    \item Expansions proposed -- 4850$^{\prime}$ (BO $\sim$ FY27) and 7400$^{\prime}$
    \item Oversight Agency -- US DOE
    \item AHJ: South Dakota Office of Risk Management
    \item Dedicated Science Facility and Education and Outreach Center
    \item In-house Safety Training Program
\vspace{-0.45em}
\end{itemize}

\paragraph{Environment:}
\begin{itemize}
\vspace{-0.45em}
\setlength\itemsep{-0.25em}
    \item Depth and Shielding 4300 mwe
    \item 5.3e-5 muons/m$^2$/s
    \item Temperature 24/28°C
    \item RH 42/79\%
    \item Rn 300 Bq/m$^3$
    \item Neutrons 1.7e-2 n/m$^2$/s
    \item Gammas 1.9 $\gamma$/cm$^2$/s
\vspace{-0.45em}
\end{itemize}

\paragraph{Space(s):} \textcolor{blue}{Blue} = available in next 10 years, \textcolor{green}{Green} = available now
\begin{itemize}
\vspace{-0.45em}
\setlength\itemsep{-0.25em}
    \item Ross Campus Clean Room -- 12.1 $\times$ 6.1 $\times$ 2.4 m$^3$
    \item Ross Campus Hall -- 30 $\times$ 3 $\times$ 2.8  m$^3$
    \item \textcolor{blue}{Davis Detector Room} -- 11 $\times$ (9.8 - 12.8) $\times$ 2.7 m$^3$
    \item Davis Machine Shop -- 9.8 $\times$ 5.3 $\times$ 2.7 m$^3$
    \item \textcolor{blue}{Davis Assay Room} -- 7.3 $\times$ 5.6 $\times$ 2.7 m$^3$
    \item Davis E-forming Room -- 6.3 $\times$ 8.7 $\times$ 2.7 m$^3$
    \item \textcolor{blue}{Davis Campus} -- 17 $\times$ 10 $\times$ 12 m$^3$
    \item \textcolor{green}{Surface Lab} -- 6.6 $\times$ 5.6 $\times$ 2.7 m$^3$
    \item \textcolor{green}{Surface Lab Rn-reduced} -- 6.6 $\times$ 8.4 $\times$ 3.3 m$^3$
\vspace{-0.45em}
\end{itemize}

\paragraph{Utilities:}
\begin{itemize}
\vspace{-0.45em}
\setlength\itemsep{-0.25em}
    \item Electrical Power -- 20,000 kW (15,000 kW available in FY27) 3 diesel generators backup 
    \item Chilled Water -- 246 kW (70 kW available)
    \item Compressed air
    \item 20 Gbps, 100 Gbps planned 
\vspace{-0.45em}
\end{itemize}

\paragraph{Access:}
\begin{itemize}
\vspace{-0.45em}
\setlength\itemsep{-0.25em}
    \item Yates Shaft -- 1.39 $\times$ 3.77 $\times$ 2.58 m, 4.8 tonnes
    \item Ross Shaft -- 1.40 $\times$ 3.70 $\times$ 3.62 m, 5.897 tonnes
    \item Slung loads -- up to 10 m
    \item Rail capacity -- 5.9 tonnes cars
\vspace{-0.45em}
\end{itemize}

\paragraph{Occupancy \& Assembly Support:}
\begin{itemize}
\vspace{-0.45em}
\setlength\itemsep{-0.25em}
    \item Multiple Clean Rooms
    \item Rail Truck (flat and lowboy)
    \item Low Background Assay (8 HPGe)
    \item Underground Refuge 144 $\times$ 4 days (250 $\times$ 4 planned) 
    \item Peak 97 people/shift, Steady state 67/shift 
    \item Multiple Cranes 2.7 tonnes
\vspace{-0.45em}
\end{itemize}

\paragraph{Surface Facilities:}
\begin{itemize}
\vspace{-0.45em}
\setlength\itemsep{-0.25em}
    \item Limited Office Space 
    \item Surface Assembly Lab 
    \item Assembly Shop
    \item Machine Shop
    \item Storage and Warehouse
\vspace{-0.45em}
\end{itemize}

\paragraph{Staff:}
8 Engineering Staff, 6 Scientists, 2 EH\&S, 178 total Staff (16 for science) modest additions planned.
\vspace{-0.45em}

\paragraph{UF Analysis:}

SURF has a remarkable record of installing experiments and obtaining world-leading results since it opened in 2009. SURF is currently fully occupied. By the end of this decade modest space will be available by reopening space at the Ross Campus (temporarily closed for DUNE excavations), the end of MJD and LZ. The SDSTA is investigating non-federal funding to develop laboratory modules capable of supporting a variety of next generation physics efforts. These would be one or two $\sim$100m modules at the 4850 Level. Concepts for a deep module at the 7400 have been renewed. If the 4850 general use module are appropriately coupled to DUNE excavation significant cost savings may be realized through mobilization and demobilization savings (O \$15M)

The space opening up from existing space could support several small R\&D efforts and modest-scaled physics efforts including first generation low-mass Dark Matter searches. 

The new laboratory module would be required to support a G3 Dark Matter experiment and/or tonne scale neutrinoless double beta decay. Beneficial occupancy was estimated by 2027.

%
% SNOLAB
%
\subsection{SNOLAB (Canada)}
\vspace{-0.45em}

\begin{figure}[ht!]
    \centering
    \includegraphics[width=0.75\textwidth]{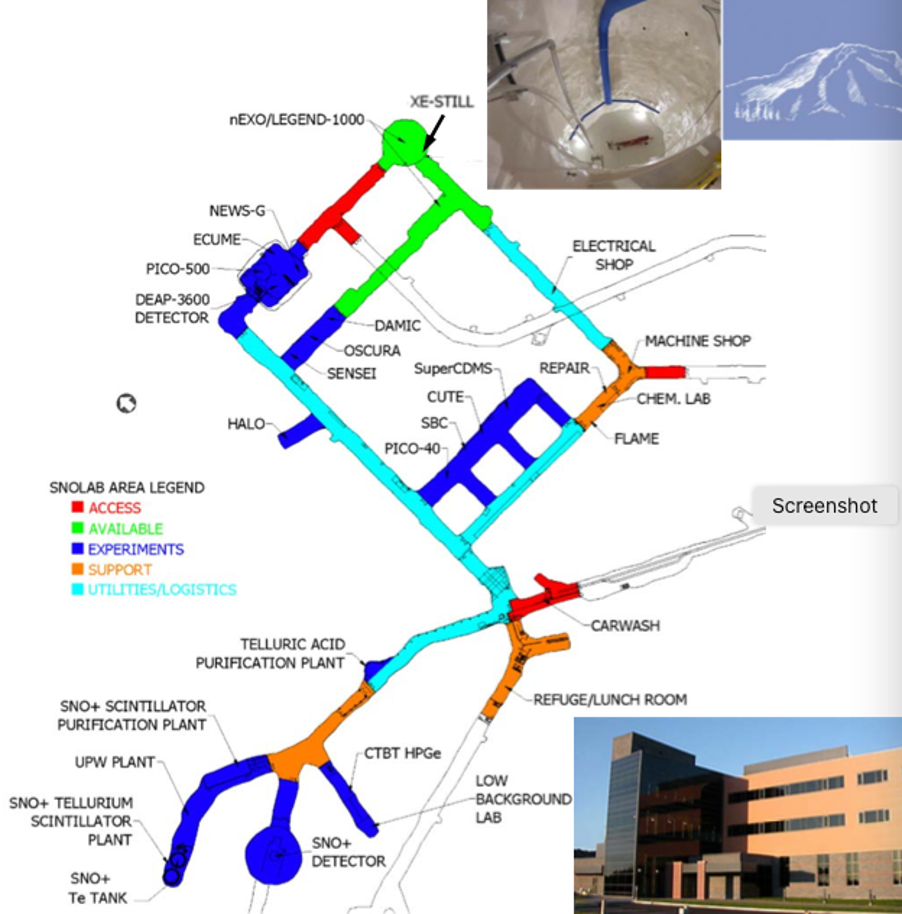}
    \caption{Map of SNOLAB.}
    \label{fig:SNOLAB}
\end{figure}

\paragraph{Description:}
\begin{itemize}
\vspace{-0.45em}
\setlength\itemsep{-0.25em}
    \item Active Nickel Mine -- 6800$^{\prime}$ 
    \item Past Experiments -- SNO, COUPP, DEAP1, DEAP-3600, miniCLEAN, NEWS-DM, PICASSO, PICO-2L, PICO-60
    \item Current Program -- SNO+(LS), PICO 40, CUTE, SENSEI, OSCURA, DAMIC, HALO
    \item Constructing -- SuperCDMS SNOLAB, DEAP-3600 II, PICO-500, ECUME, NEWS-G, SNO+(Te), SBC, 
    \item Expansions proposed -- 6800 (time frame is uncertain) 
    \item Oversight Agencies -- CFI, NSERC, Ontario Opportunities Fund
    \item AHJ: N/A (Suspect it is Vale and Canadian Ministry of Mines)
    \item Shared with Vale Mining 
    \item Vale Training + SNOLAB Specific
\vspace{-0.45em}
\end{itemize}

\paragraph{Environment:}
\begin{itemize}
\vspace{-0.45em}
\setlength\itemsep{-0.25em}
    \item Depth and Shielding 6000 mwe, $<$ 0.27 muons/m$^2$/day
    \item Temperature 18/23°C
    \item RH 30/50\%
    \item Rn 100 Bq/m$^3$
    \item Neutrons 4.1e3 n/m$^2$/day
    \item Gammas 510 $\gamma$/m$^2$/d
\vspace{-0.45em}
\end{itemize}

\paragraph{Space(s):} \textcolor{blue}{Blue} = available in next 10 years, \textcolor{green}{Green} = available now
\begin{itemize}
\vspace{-0.45em}
\setlength\itemsep{-0.25em}
    \item SNO Cavity 22m dia $\times$ 35 m tall -- SNO+
    \item Ladder Labs -- Pico-40, SBC, CUTE, SuperCDMS, HALO, SENSEI, OSCURA, DAMIC, NEWS-G, additional labs
    \item \textcolor{blue}{Cube Hall} -- DEAP-3600, ECUME, PICO-500
    \item \textcolor{green}{Cryo-pit} -- tonne scale DBD
\vspace{-0.45em}
\end{itemize}

\paragraph{Utilities:}
\begin{itemize}
\vspace{-0.45em}
\setlength\itemsep{-0.25em}
    \item Electrical Power 3 MW 
    \item Standby Generator 3MW
    \item Chilled Water 320 tons
    \item Waste heat 1.5 MW
    \item 150 CFM compressed air
    \item 10  Gbps
\vspace{-0.45em}
\end{itemize}

\paragraph{Access:}
\begin{itemize}
\vspace{-0.45em}
\setlength\itemsep{-0.25em}
    \item Cage 54” $\times$ 144” $\times$ 68” 5500 lbs
    \item 2 $\times$ 10-hour shifts/day, 5 days/week
    \item Mine Shutdown 2 weeks/year
    \item Shaft work 2023-2025 to limit u/g access
\vspace{-0.45em}
\end{itemize}

\paragraph{Occupancy \& Assembly Support:}
\begin{itemize}
\vspace{-0.45em}
\setlength\itemsep{-0.25em}
    \item Laboratory Clean Room Class 2000
    \item Rail Truck
    \item Low Background Assay 
    \item Peak 120 people/shift, Steady state 50/shift 
    \item Multiple Cranes: 2., 2., 10.  Tonnes
\vspace{-0.45em}
\end{itemize}

\paragraph{Surface Facilities:}
\begin{itemize}
\vspace{-0.45em}
\setlength\itemsep{-0.25em}
    \item Significant Office Space 34k sf 
    \item 4700 sf clean room labs 
    \item Assembly Shop
    \item Machine Shop
    \item Storage and Warehouse
    \item Chemistry Labs
\vspace{-0.45em}
\end{itemize}

\paragraph{Staff:}
60 Technical, 40 Science, 5 EH\&S, 20 Admin
\vspace{-0.45em}

\paragraph{UF Analysis:}
SNOLAB has a substantial science program primarily focused on Dark Matter and neutrinoless double-beta decay.  The SNO detector has been undergoing upgrades since 2006 to host  Te-based neutrinoless double-beta decay experiment SNO+.  SuperCDMS at SNOLAB is in their 2nd year of integration and installation. The $\sim$20m cryopit is reserved for tonne-scale neutrinoless double beta decay.  Modest floor space is currently available and anticipated in the coming years suitable for R\&D and smaller-scale experimental efforts. The cube hall could be made available in the coming several years to host a G3-DM experiment. Expansion plans at the 6800 level have been presented.  

%
% BOULBY
%
\subsection{Boulby (UK)}
\vspace{-0.45em}

\begin{figure}[ht!]
    \centering
    \includegraphics[width=0.75\textwidth]{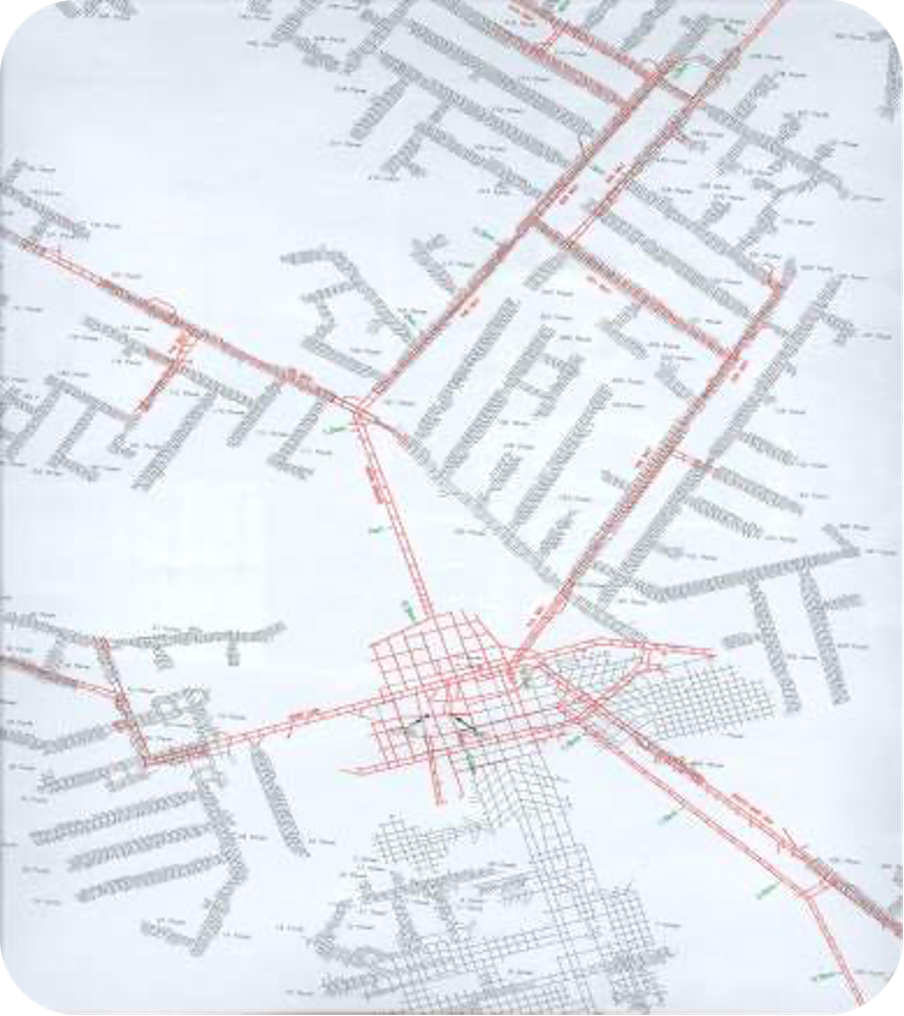}
    \caption{Map of Boulby.}
    \label{fig:Boulby}
\end{figure}

\paragraph{Description:}
\begin{itemize}
\vspace{-0.45em}
\setlength\itemsep{-0.25em}
    \item Active Potash Mine -- 2850 mwe 
    \item Past Experiments -- ZEPLIN, DRIFT, NAIAD
    \item Current Program -- CYGNUS, Low Background Assay, NEWS-G/Dark Sphere R\&D
    \item Planning -- DarkSPHERE, LXe-G3
    \item Expansions proposed -- 25m $\times$ 25m tall cavity, 14,000 m$^3$ Rooms
    \item Oversight Agencies -- STFC/UKRI
    \item AHJ: STFC/ICL-UK
    \item Shared with ICL-UK
    \item Safety STFC Site Specific
\vspace{-0.45em}
\end{itemize}

\paragraph{Environment:}
\begin{itemize}
\vspace{-0.45em}
\setlength\itemsep{-0.25em}
    \item Depth and Shielding 2850 mwe -- 4e-8/cm$^2$/s
    \item Temperature 21°C
    \item RH 40/50\%
    \item Rn 2.5 Bq/m$^3$
    \item Neutrons 4.1 e-6 n/cm$^2$/s
    \item Gammas 0.13 $\gamma$/cm$^2$/s
\vspace{-0.45em}
\end{itemize}

\paragraph{Space(s):} \textcolor{blue}{Blue} = available in next 10 years, \textcolor{green}{Green} = available now
\begin{itemize}
\vspace{-0.45em}
\setlength\itemsep{-0.25em}
    \item Main Hall -- 60 $\times$ 6 $\times$ 3.8 m 10k CR
    \item BUGS -- 20 $\times$ 6 $\times$ 3.8 1k CR
    \item LEC -- 10 $\times$ 6 $\times$ 6 10k CR
    \item Outside -- 20 $\times$ 6 $\times$ 3.8
    \item Discussion of Major New Excavations -- 25m dia $\times$ 25 m tall
\vspace{-0.45em}
\end{itemize}

\paragraph{Utilities:}
\begin{itemize}
\vspace{-0.45em}
\setlength\itemsep{-0.25em}
    \item Electrical Power 100 kW 
    \item Standby Generator N/A
    \item Chilled Water N/A
    \item Waste heat N/A
    \item Compressed air Yes
    \item 2 $\times$ 40 Gb to surface 1 Gb offsite
\vspace{-0.45em}
\end{itemize}

\paragraph{Access:}
\begin{itemize}
\vspace{-0.45em}
\setlength\itemsep{-0.25em}
    \item Cage 2m $\times$ 2m  5500 lbs
    \item 3 shifts/day
    \item 15 science/cage, 3-4 cages / shift
    \item Peak 30, Steady-state 10, Refuge 30, expected to increase to 60/50/100
\vspace{-0.45em}
\end{itemize}

\paragraph{Occupancy \& Assembly Support:}
\begin{itemize}
\vspace{-0.45em}
\setlength\itemsep{-0.25em}
    \item 4000m$^3$ Class 10k, 1800m$^3$ Class 1k
    \item Rn-reduced Air
    \item Low Background Assay 
    \item Peak 120 people/shift, Steady state 50/shift 
    \item Multiple Cranes: 5, 10 tonnes
\vspace{-0.45em}
\end{itemize}

\paragraph{Surface Facilities:}
\begin{itemize}
\vspace{-0.45em}
\setlength\itemsep{-0.25em}
    \item Some Office Space, planning for 2600m$^2$
    \item Assembly Shop - N/A
    \item Machine Shop - N/A
    \item Storage and Warehouse
    \item Chemistry Labs- N/A
\vspace{-0.45em}
\end{itemize}

\paragraph{Staff:}
5 Technical, 5 Science, 2 EH\&S, 2 Admin
\vspace{-0.45em}

\paragraph{UF Analysis:}
Boulby is a research facility in northern England hosted by a potash mine.  It has hosted a variety of dark matter experiments in the past including the UK ZEPLIN efforts. There is a modest staff to assist experimental efforts.  Their low background assay capabilities are very significant and expert.  Existing facilities support modest R\&D efforts with some small capacity to expand this decade.  Notably mining in the salt domes is relatively inexpensive and expeditious.  They presented ideas for an expansion to host a G3 dark matter experiment at the existing level or at a lower level. 

%
% KAMIOKA
%
\subsection{Kamioka (Japan)}
\vspace{-0.45em}

\begin{figure}[ht!]
    \centering
    \includegraphics[width=0.75\textwidth]{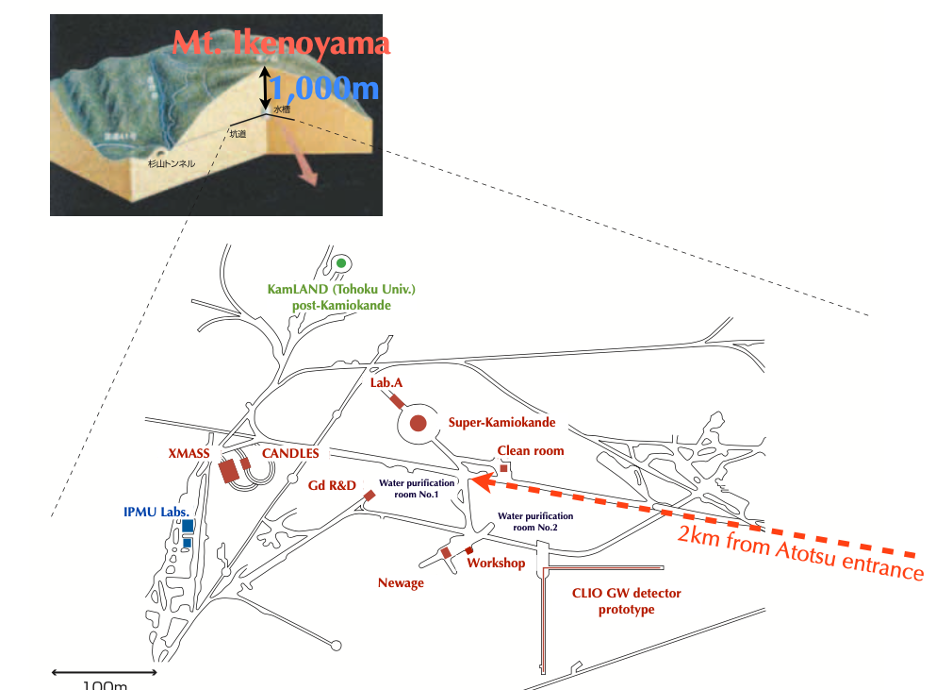}
    \caption{Map of Kamioka.}
    \label{fig:Kamioka}
\end{figure}

\paragraph{Description:}
\begin{itemize}
\vspace{-0.45em}
\setlength\itemsep{-0.25em}
    \item Former Mine -- 2700 mwe 
    \item Horizontal Access -- roadway
    \item Past Experiments -- KamiokaNDE I, II, III, K2K, XMASS, KamLAND 
    \item Current Program -- SuperK, T2K, KamLAND-Zen, CLIO, KAGRA, NEWAGE, CANDLES
    \item Constructing -- Hyper-Kamiokande
    \item Oversight Agencies -- MEXT
    \item AHJ: Labour Standard Inspection Office
    \item Shared with Kamioka Mining \& Smelting
    \item Mine Safety Training, University Safety Training
\vspace{-0.45em}
\end{itemize}

\paragraph{Environment:}
\begin{itemize}
\vspace{-0.45em}
\setlength\itemsep{-0.25em}
    \item Depth and Shielding 2700 mwe -- 1.5e-7/cm$^2$/s with mean energy of 273 GeV
    \item Temperature 13°C
    \item RH 93\%
    \item Rn 64 Bq/m$^3$
    \item Thermal Neutrons 12.5 e-5 n/cm$^2$/s
    \item Gammas 2.7e-6 $\gamma$/cm$^2$/sr/s
\vspace{-0.45em}
\end{itemize}

\paragraph{Space(s):} \textcolor{blue}{Blue} = available in next 10 years, \textcolor{green}{Green} = available now
\begin{itemize}
\vspace{-0.45em}
\setlength\itemsep{-0.25em}
    \item Super-K -- 39.3 dia $\times$ 41.4 m
    \item Hyper-K -- 68 dia $\times$ 71 m
    \item KamLAND -- 29 dia $\times$ 32m
    \item Lab C -- 20 $\times$ 15 $\times$ 15
    \item Lab D -- 14 $\times$ 7 $\times$ 11
    \item Lab E -- 15 $\times$ 10 $\times$ 9
    \item Lab G -- 50 $\times$ 8 $\times$ 7
\vspace{-0.45em}
\end{itemize}

\paragraph{Utilities:}
\begin{itemize}
\vspace{-0.45em}
\setlength\itemsep{-0.25em}
    \item Electrical Power -- N/A 
    \item Standby Generator -- N/A
    \item Chilled Water -- N/A
    \item Waste heat -- N/A
    \item Compressed air -- N/A
    \item XX Gb 
    \item Surface air $\sim$70 Bq/m$^3$
    \item Rn-free air 36m$^3$/hour
\vspace{-0.45em}
\end{itemize}

\paragraph{Access:}
\begin{itemize}
\vspace{-0.45em}
\setlength\itemsep{-0.25em}
    \item Horizontal Access 
    \item Open 24/7/365 
    \item Road -- 3.2m tall $\times$ 4.5 wide
    \item Rail -- 2.5m $\times$ 3
    \item 10 tonne truck
    \item Peak 100, Steady-state 2, Refuge N/A
\vspace{-0.45em}
\end{itemize}

\paragraph{Occupancy \& Assembly Support:}
\begin{itemize}
\vspace{-0.45em}
\setlength\itemsep{-0.25em}
    \item 100m$^2$ room Class 10, 50m$^2$ Class 200k
    \item Rn-reduced Air, Rn-free Air
    \item Low Background Assay 
    \item Peak 100 people/shift, Steady state 2/shift 
    \item Multiple Cranes: 1, 2 tonnes
\vspace{-0.45em}
\end{itemize}

\paragraph{Surface Facilities:}
\begin{itemize}
\vspace{-0.45em}
\setlength\itemsep{-0.25em}
    \item Office Space for 50
    \item 6 Meeting Rooms 
    \item Assembly Shop N/A
    \item Machine Shop N/A
    \item Storage and Warehouse
    \item Multiple Surface Labs 
    \item Low Background Assay
\vspace{-0.45em}
\end{itemize}

\paragraph{Staff:}
3 Technical, 26 Science, 0 EH\&S, 8 Admin
\vspace{-0.45em}

\paragraph{UF Analysis:}
The venerable Kamioka laboratory continues to expand its scientific portfolio from the original proton decay, to atmospheric neutrinos, solar neutrinos, long baseline neutrinos, neutrinoless double-beta decay, and recently several gravity wave arrays. They have indicated some modest rooms available in the near future for R\&D and smaller scale experiments. The Super-K cavity may also become available this decade. A major expansion is underway to host the Hyper-K experiment in an adjacent mountain.

%
% LNGS
%
\subsection{LNGS (Italy)}
\vspace{-0.45em}

\begin{figure}[ht!]
    \centering
    \includegraphics[width=0.75\textwidth]{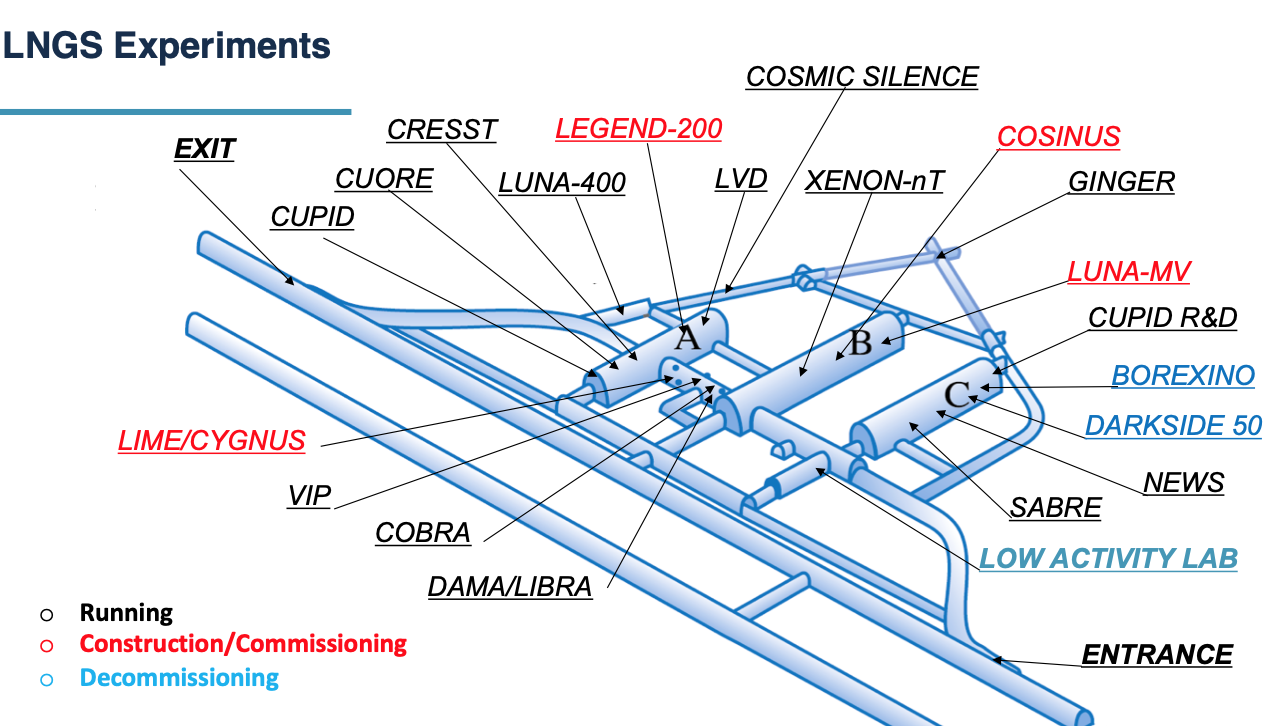}
    \caption{Map of LNGS with experiments and locations identified.}
    \label{fig:LNGS}
\end{figure}

Note: Information gleaned from the literature, no response to the questionnaire was received by UF.
\vspace{-0.45em}

\paragraph{Description:}
\begin{itemize}
\vspace{-0.45em}
\setlength\itemsep{-0.25em}
    \item Dedicated Physics Laboratory -- 1400 m vertical depth 
    \item Horizontal Access (Highway Tunnel)
    \item Past Experiments -- Borexino, OPERA, MACRO, GALLEX, Icarus, Warp, GERDA
    \item Current Program -- COBRA, COSINUS, CRESST, CUORE, CUPID-0, CUPID, DAMA, DARKSIDE, ERMES, LEGEND-200, GINGER, LUNA, LVD, NEWS-DM,  SABRE, VIP, XENON
    \item Operations Agency -- INFN
    \item Separate Environmental oversight
    \item Restrictions on chemical use u/g
\vspace{-0.45em}
\end{itemize}

\paragraph{Environment:}
\begin{itemize}
\vspace{-0.45em}
\setlength\itemsep{-0.25em}
    \item Depth and Shielding 3800 mwe and effective 2800 mwe -- $\sim$3.4e-8/cm$^2$/s 
    \item Temperature 15-26°C
    \item RH 30-90\%
    \item Rn 20-80 Bq/m$^3$
    \item Thermal Neutrons $<$4.6e-6 n/cm$^2$/s
    \item Gammas 0.3-0.4 $\gamma$/cm$^2$/s
\vspace{-0.45em}
\end{itemize}

\paragraph{Space(s):} \textcolor{blue}{Blue} = available in next 10 years, \textcolor{green}{Green} = available now
\begin{itemize}
\vspace{-0.45em}
\setlength\itemsep{-0.25em}
    \item 3 halls each 100 $\times$ 20 $\times$ 18 m
    \item Total volume 180,000 m$^3$
    \item Retirement of experiments in \textcolor{blue}{Halls A and C} will make space available for new experiments such as a G3 dark matter experiment.
\vspace{-0.45em}
\end{itemize}

\paragraph{Utilities:}
\begin{itemize}
\vspace{-0.45em}
\setlength\itemsep{-0.25em}
    \item Electrical Power -- N/A
    \item Standby Generator -- N/A
    \item Chilled Water -- N/A
    \item Waste heat -- N/A
    \item Compressed air -- N/A
    \item N/A  Gbps 
    \item Rn 20-80 Bq/m$^3$
\vspace{-0.45em}
\end{itemize}

\paragraph{Access:}
\begin{itemize}
\vspace{-0.45em}
\setlength\itemsep{-0.25em}
    \item Highway Tunnel
    \item 20 tonnes (estimate)
    \item Access 24/7
    \item N/A people/shift peak
\vspace{-0.45em}
\end{itemize}

\paragraph{Occupancy \& Assembly Support:}
\begin{itemize}
\vspace{-0.45em}
\setlength\itemsep{-0.25em}
    \item No common CR
    \item Low Background Assay 
    \item 225 people average
    \item Cranes: 5  tonnes
\vspace{-0.45em}
\end{itemize}

\paragraph{Surface Facilities:}
\begin{itemize}
\vspace{-0.45em}
\setlength\itemsep{-0.25em}
    \item Substantial Surface Support 17000 m$^2$ 
    \item Chemistry Lab
    \item ULB Lab
\vspace{-0.45em}
\end{itemize}

\paragraph{Staff:}
110 staff personnel, 14 researches, 35 engineers, and 38 technicians.
\vspace{-0.45em}

\paragraph{UF Analysis:}
No response was obtained from LNGS - Gran Sasso. This information was culled from the literature. LNGS has an impressive array of physics experiments and excellent support of underground science developed over the past 40 years. The removal of Borexino and LVD has opened space in Gran Sasso. In Hall C, there is space for three large cryogenic experiments (\textit{e.g.,} Legend-1000, DarkSide-20k, and a G3 dark matter experiment).

%
% YEMILAB
%
\subsection{Yemilab (Korea)}
\vspace{-0.45em}

\begin{figure}[ht!]
    \centering
    \includegraphics[width=0.75\textwidth]{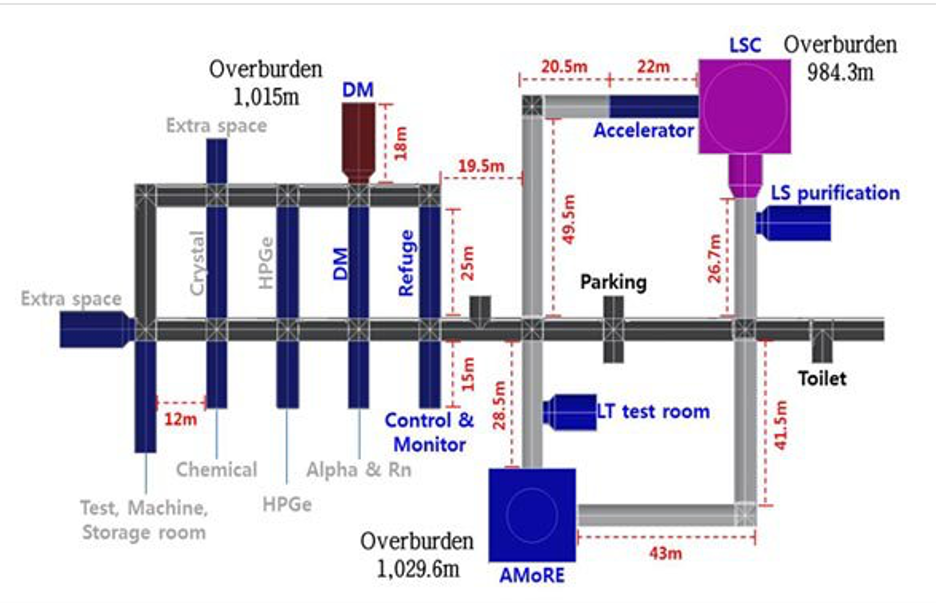}
    \caption{Map of YemiLab.}
    \label{fig:YemiLab}
\end{figure}

\paragraph{Description:}
\begin{itemize}
\vspace{-0.45em}
\setlength\itemsep{-0.25em}
    \item Active Mine -- 2500 mwe 
    \item Vertical Access, Dedicated Access Drift
    \item Past Experiments -- none, new laboratory! 
    \item Current Program -- Establishing Research Program
    \item Constructing -- COSINE, AMoRE, LSC,  IsoDAR, KNU, KIGAM, Low Background  Counting
    \item Operations Agency -- Center for Underground Physics in IBS
    \item Oversight Agencies -- Ministry of Science and ICT 
    \item Shared with SM Handeok Iron-Mine \& Construction
    \item Mine Safety Training, Yemilab Safety Training
\vspace{-0.45em}
\end{itemize}

\paragraph{Environment:}
\begin{itemize}
\vspace{-0.45em}
\setlength\itemsep{-0.25em}
    \item Depth and Shielding 2500 mwe -- TBD $\sim$ 1.e-6/cm$^2$/s 
    \item Temperature 25°C
    \item RH TBD\%
    \item Rn 20-2000 Bq/m$^3$
    \item Thermal Neutrons TBD n/cm$^2$/s
    \item Gammas 516 $\gamma$/s
\vspace{-0.45em}
\end{itemize}

\paragraph{Space(s):} \textcolor{blue}{Blue} = available in next 10 years, \textcolor{green}{Green} = available now
\begin{itemize}
\vspace{-0.45em}
\setlength\itemsep{-0.25em}
    \item Operations -- 15 $\times$ 5 $\times$ 5 m
    \item LSC -- 22 $\times$ 22 $\times$ 8 + 20 dia $\times$ 20 h
    \item Purification -- 15 $\times$ 7 $\times$ 7
    \item AMoRE -- 21 $\times$ 21 $\times$ 16 + 13 $\times$ 8 $\times$ 8
    \item COSINE -- 15 $\times$ 15 $\times$ 8 + 25 $\times$ 5 $\times$ 5
    \item KNU -- 15 $\times$ 8 $\times$ 8
    \item KIGAM -- 15 $\times$ 8 $\times$ 8
    \item IsoDAR -- 15 $\times$ 12 $\times$ 10, 20 $\times$ 7 $\times$ 7
    \item \textcolor{green}{Open} -- 25 $\times$ 5 $\times$ 5, 3 $\times$(12 $\times$ 5 $\times$ 5)
\vspace{-0.45em}
\end{itemize}

\paragraph{Utilities:}
\begin{itemize}
\vspace{-0.45em}
\setlength\itemsep{-0.25em}
    \item Electrical Power -- 1800kW 
    \item Standby Generator -- 1800kW
    \item Chilled Water -- 15 kW (35kW by 2023)
    \item Waste heat -- 180 kW
    \item Compressed air -- N/A
    \item 10 Gbps 
    \item Rn -- TBD $\sim$ 20-2000 Bq/m$^3$
\vspace{-0.45em}
\end{itemize}

\paragraph{Access:}
\begin{itemize}
\vspace{-0.45em}
\setlength\itemsep{-0.25em}
    \item Vertical Access -- 2.8 $\times$ 1.5 $\times$ 3.5 m:  1.5 tonnes, There is also a ramp system for trucks shared with mining.
    \item Access TBD 3 shift/day 24 hr monitoring
    \item 40 people/shift peak
    \item 40 people refuge
\vspace{-0.45em}
\end{itemize}

\paragraph{Occupancy \& Assembly Support:}
\begin{itemize}
\vspace{-0.45em}
\setlength\itemsep{-0.25em}
    \item No common CR
    \item Rn-reduced Air 50 CMH
    \item Low Background Assay 
    \item Peak 40 people/shift, Steady state TBD/shift 
    \item Multiple Cranes: 5  tonnes
\vspace{-0.45em}
\end{itemize}

\paragraph{Surface Facilities:}
\begin{itemize}
\vspace{-0.45em}
\setlength\itemsep{-0.25em}
    \item 2500 m$^2$ Office Space
    \item 64 m$^2$ Clean Room
    \item 64 m$^2$ Assembly Shop 
    \item 32 m$^2$ Machine Shop 
    \item 64 m$^2$ Storage and Warehouse
    \item 64 m$^2$ Chemistry Labs
\vspace{-0.45em}
\end{itemize}

\paragraph{Staff:}
7 Technical, 5 Science Expansion anticipated to: 10 Technical, 8 Science  1 EH\&S, 2 Admin
\vspace{-0.45em}

\paragraph{UF Analysis:}
Yemilab opened in 2022 with a new underground laboratory, significantly upgrading the shielding from the earlier Y2L facility further north in Korea.  The design of the facility has clearly applied lessons from earlier experiments.  Nearly all the new rooms are filled with an initial suite of experiments.

\subsection{General Conclusions on Availability of Underground Space in the Coming Decade}

Currently the underground facilities that reported to the Underground Facilities are heavily subscribed. Toward the end of this decade several facilities indicated they anticipate additional space becoming available as the current experiments complete their programs.  SNOLAB has reserved their cryopit for tonne-scale neutrinoless double-beta decay efforts, otherwise the space and resources current available or becoming available is appropriate for R\&D efforts and several smaller scale experiments, e.g. the scale of Generation-2 Dark Matter efforts currently underway, or Neutrinoless double-beta decay experiments of O(100kgs).  No facility responding to the survey currently has the space or capacity to host a Generation-3 Dark Matter experiment.

\section{Recommendations from 2014 P5 Report}

There are several recommendations from the 2014 P5 report of importance for the 2022 Snowmass planning. These are provided here, within the context of the 2014 P5 Report. The first relates to LBNF and DUNE.

\begin{quote}

\textcolor{blue}{Neutrino Oscillation Experiments}

Short- and long-baseline oscillation experiments directly probe three of the questions of the neutrino science Driver: How are the neutrino masses ordered? Do neutrinos and antineutrinos oscillate differently? Are there additional neutrino types and interactions? There is a vibrant international neutrino community invested in pursuing the physics of neutrino oscillations. The U.S. has unique accelerator capabilities at Fermilab to provide neutrino beams for both short- and long-baseline experiments, with some experiments underway. A long-baseline site is also available at the Sanford Underground Research Facility in South Dakota. Many of these current and future experiments and projects share the same technical challenges. Interest and expertise in neutrino physics and detector development of groups from around the world combined with the opportunities for experiments at Fermilab provide the essentials for an international neutrino program.

\textit{[Recommendation 12 is made at this point in the P5 report. The following text provides details on the requirements to meet the science objectives of LBNF and DUNE. The discussion then progresses to requirements for underground facilities\ldots]}

Key preparatory activities will converge over the next few years: in addition to the international reformulation described above, PIP-II design and project definition will be nearing completion, as will the necessary refurbishments to the Sanford Underground Research Facility. Together, these will set the stage for the facility to move from the preparatory to the construction phase around 2018. The peak in LBNF construction will occur after HL-LHC peak construction.

\textbf{Recommendation 13: Form a new international collaboration to design and execute a highly capable Long-Baseline Neutrino Facility (LBNF) hosted by the U.S. To proceed, a project plan and identified resources must exist to meet the minimum requirements in the text. LBNF is the highest-priority large project in its timeframe.}

\end{quote}

The second and third 2014 P5 Recommendations related to underground facilities are focused on direct detection of particle like dark matter.

\begin{quote}

\textcolor{blue}{Dark Matter}

The experimental challenge of discovery and characterization of dark matter interactions with ordinary matter requires a multi-generational suite of progressively more sensitive and ambitious direct detection experiments. This is a highly competitive, rapidly evolving field with excellent potential for discovery. The second-generation direct detection experiments are ready to be designed and built, and should include the search for axions, and the search for low-mass (<10 GeV) and high-mass WIMPs. Several experiments are needed using multiple target materials to search the available spin-independent and spin-dependent parameter space. This suite of experiments should have substantial cross-section reach, as well as the ability to confirm or refute current anomalous results. Investment at a level substantially larger than that called for in the 2012 joint agency announcement of opportunity will be required for a program of this breadth.

\textbf{Recommendation 19: Proceed immediately with a broad second-generation (G2) dark matter direct detection program with capabilities described in the text. Invest in this program at a level significantly above that called for in the 2012 joint agency announcement of opportunity.}

The results of G2 direct detection experiments and other contemporaneous dark matter searches will guide the technology and design of third-generation experiments. As the scale of these experiments grows to increase sensitivity, the experimental challenge of direct detection will still require complementary experimental techniques, and international cooperation will be warranted. The U.S. should host at least one of the third-generation experiments in this complementary global suite.

\textbf{Recommendation 20: Support one or more third-generation (G3) direct detection experiments, guided by the results of the preceding searches. Seek a globally complementary program and increased international partnership in G3 experiments.}

\end{quote}

\subsection{Discussion of 2014 P5 Recommendations}

The LBNF/DUNE program is of great importance for Underground Facilities domestically and internationally.  Excavations at SURF to house Phase 1 and 2 of the DUNE detectors are making good progress, with farsite excavations passing the 35\% completion point.  As will be discussed further below, during the Snowmass meeting in Seattle (July 2022) some in the community noted the DUNE excavations and DUNE cavities provide opportunities for cost-effective expansion of underground facilities and options to potentially host experiments at SURF in the coming years.

Good progress has been made by the US Generation-2 Dark Matter experiments with LZ reporting world-leading results and beginning its full science mission.  ADMX has updated their searches continually since 2014 and is currently being upgraded.  SuperCDMS at SNOLAB continues its integration and installation. LZ and ADMX are providing information to assist with the design of next generation direct detection experiments as presented in Recommendation 19, above.

There was active discussion on next-generation direction detection experiments at the Snowmass meeting in Seattle including several efforts with strong U.S. participation. In our survey of facilities there exist no obvious site to host a Generation-3 experiment. Three sites presented proposals to produce new space to host such an experiment: Boulby, SNOLAB, and SURF. 

Dark Matter dominated much of the discussion in Seattle across multiple frontiers.  Several collaborations and consortia presented plans for Generation-3 experiments. These experiments are actively applying technology advances from the suite of Generation-2 experiments and a very broad expansion of R\&D efforts. As discussed further below, much of the discussion at the Snowmass meeting in Seattle revolved around how to move these programs forward with continuity of U.S. leadership.

\section{In-person Discussions at Snowmass in Seattle}

There were three notable discussion areas reflecting on the ability of Underground Facilities to advance high energy physics experiments in the coming decades.

1. \textit{Consideration of utilizing unoccupied space created by LBNF excavations to temporarily host experiments.} LBNF will create four cavities at the 4850 Level of SURF.  Phase 1 of DUNE would use two of these.  The other two would remain vacant until Phase 2 was defined and the project established.  With the encouragement of Neutrino Frontier liaisons, Underground Facilities solicited concrete letters of interest to help evaluate using these cavities, resulting several short descriptions being provided.  These included several smaller scale R\&D efforts and at least one Generation-3 Dark Matter experiment. There was clear conscientious awareness of the potential for timing conflicts with DUNE Phase 2.

2. \textit{Designating the Sanford Underground Research Facility as a DOE User Facility.} When DOE assumed oversight of Sanford Lab in 2011, it was managed by Berkeley Lab for DOE, noting the familiarity of the facility by Berkeley Lab.  As LBNF/DUNE efforts at the far site grew, management was passed to Fermilab, as these far site efforts drove a predominance of scientific efforts at SURF.  More recently the South Dakota Science and Technology Authority (SDSTA) obtained a Cooperative Agreement with DOE to facilitate operations and construction activities. During the Snowmass meeting in Seattle UF learned of efforts by SDSTA to elevate SURF from a research facility to a DOE Office of Science User Facility.

3. \textit{Creation of significant new laboratory space at the 4850 Level of SURF.}  The SDSTA management presented plans during the Snowmass process for advancing new excavations at the 4850L. The SDSTA is pursuing State of South Dakota and Private funds for this excavation. Initial plans have identified two locations near the Ross shaft that are each suitable for one or two 100m long laboratory modules.  A single 100m module could support many designs for a Generation-3 dark matter experiment and a tonne-scale neutrinoless double-beta decay experiment, for example. A notable consideration is that scheduling the new excavation in concert with LBNF could save significant costs by reducing or eliminating mobilization and demobilization costs. These savings were estimated at O(\$15M) based on analogous costs being incurred during the LBNF excavation program.

\section{Conclusions}

UF liaised with most of the other Frontiers to assess the needs for underground space in the coming decades.  Requests for significant R\&D and prototype space were received.  The Generation-3 dark matter experimentalists expressed need for a significant underground hall to fulfill the 2014 P5 recommendation at a domestic underground location.  Existing underground facilities are able to satisfy some of the smaller space requests internationally, but none in the U.S. The existing underground laboratories do not have adequate capacity for the wide spectrum of requests presented at Snowmass.

Several emergent fields have been identified, requesting underground space, including quantum information science (QIS), quantum computing (QC), and atom interferometry. These new research efforts may lead to additional facility requirements and unique space needs in the coming decade.

At the Snowmass meeting in Seattle, options to provide a new laboratory module at SURF’s 4850 with state and private resources were presented. Expansion concepts for Boulby, also aimed at a G3 experiment, were presented.  SNOLAB presented options for further expansion. These are the three options for the G3 experiment.

In summary, high energy physics research at underground facilities has blossomed since the last Snowmass in 2013. The 2014 P5 Recommendations related to underground facilities are nearly fulfilled. The outlook for the coming decade is a U.S. high energy physics program that continually utilizes underground facilities to advance the scientific understand of neutrinos and press for the discovery of dark matter.

%%%%%%%%%%%%%%%%%%%%%%%%%%%%%%%%%%%%%%%%%%

%  If you would like to use BibTEX for the bibliography, please feel free to do so.  It is not required.

%  To use BibTeX,

%    1.  uncomment the following two lines,
%    2.  comment out everything below from  \begin{thebibliography}{99}   to \end{thebibliography).
%    3.  create the file  myreferences.bib in this directory, and process this file in the usual way

\bibliographystyle{JHEP}
\bibliography{Underground_UF06_TopicalReport}

%%%%%%%%%%%%%%%%%%%%%%%%%%%%%%%%%%%%%%%%%

%\begin{thebibliography}{99}
%\input Underground/UF06/bibliography.tex
%\end{thebibliography}

\end{document}